\documentclass{ws-rv9x6}  
\def\fsl#1{\setbox0=\hbox{$#1$}                 
   \dimen0=\wd0                                 
   \setbox1=\hbox{/} \dimen1=\wd1               
   \ifdim\dimen0>\dimen1                        
      \rlap{\hbox to \dimen0{\hfil/\hfil}}      
      #1                                        
   \else                                        
      \rlap{\hbox to \dimen1{\hfil$#1$\hfil}}   
      /                                         
   \fi}                                         %

\begin{document}

\setcounter{chapter}{0}

\chapter{SPONTANEOUS NAMBU-GOLDSTONE CURRENTS GENERATION DRIVEN 
BY MISMATCH}

\vspace{-5.6cm}

\hfill {\small TKYNT-06-3 }

\vspace{5.6cm}

\markboth{Mei Huang}{Spontaneous Nambu-Goldstone currents generation}

\author{Mei Huang}

\address{Physics Department, 
     University of Tokyo, \\
     Hongo, Bunkyo-ku, Tokyo 113-0033, Japan\\
     E-mail: huang@nt.phys.s.u-tokyo.ac.jp }

\begin{abstract}
We review recent progress of understanding and resolving instabilities 
driven by mismatch between the Fermi surfaces of the pairing quarks
in 2-flavor color superconductor. With the increase of mismatch, the 2SC 
phase exhibits chromomagnetic instability as well as color neutral baryon 
current instability. We describe the 2SC phase in the nonlinear realization 
framework,  and show that each instability indicates the spontaneous generation 
of the corresponding pseudo Nambu-Golstone current. The Nambu-Goldstone 
currents generation state covers the gluon phase as well as the one-plane wave 
LOFF state. We further point out that, when charge neutrality condition is required, 
there exists a narrow unstable LOFF (Us-LOFF) window, where not only off-diagonal 
gluons but the diagonal 8-th gluon cannot avoid  the magnetic instability. In this 
Us-LOFF window, the diagonal magnetic instability cannot be cured by off-diagonal 
gluon condensate in the color superconducting phase.   
\end{abstract}

\section{Introduction}

Studying QCD at finite baryon density is the traditional subject of nuclear 
physics. The behaviour of QCD at finite baryon density and low temperature is 
central for astrophysics to understand the structure of compact stars, and 
conditions near the core of collapsing stars (supernovae, hypernovae).
It is known that sufficiently cold and dense baryonic matter is in the color 
superconducting phase. This was proposed several decades ago by 
Frautschi \cite{Frautschi} and Barrois \cite{Barrois} by noticing that one-gluon 
exchange between two quarks is attractive in the color antitriplet channel. 
From BCS theory \cite{BCS}, we know 
that if there is a weak attractive interaction in a cold Fermi sea, the system is 
unstable with respect to the formation of particle-particle Cooper-pair condensate 
in the momentum space. Studies on color superconducting phase in 1980's can 
be found in  Ref. \cite{Bailin-Love}. 
The topic of color superconductivity stirred a lot of interest in recent years \cite{cs,cfl,weak,weak-cfl}.  
For reviews on recent progress of color superconductivity see, for example, Ref.~\cite{reviews}.

The color superconducting phase may exist in the central region of compact stars. 
To form bulk matter inside compact stars,  the charge neutrality condition as well as $\beta$ equilibrium are 
required \cite{absence2sc,neutral_steiner,neutral_huang}. This induces mismatch between the Fermi 
surfaces of the pairing quarks. It is clear that the Cooper pairing will be eventually destroyed with the increase of 
mismatch. 

Without the constraint from the charge neutrality condition, the system may exhibit a first order phase transition 
from the color superconducting phase to the normal phase when the mismatch increases \cite{asym-2sc}.
It was also found that the system can experience a spatial non-uniform LOFF (Larkin-Ovchinnikov-Fudde-Ferrell) 
state \cite{loff-orig,LOFF} in a certain window of moderate mismatch.

It is still not fully understood how the Cooper pairing will be eventually destroyed by
increasing mismatch in a charge neutral system.
The charge neutrality condition plays an essential role in determining the ground state of the neutral system.
If the charge neutrality condition is satisfied globally, and also if the surface tension is small, the mixed phase will be 
favored  \cite{mixed-phase}. It is difficult to precisely calculate the surface tension in the mixed phase, thus in the 
following, we would like to focus on the homogeneous phase when the charge neutrality condition is required locally. 
 
It was found that homogeneous neutral cold-dense quark matter can be in the gapless 2SC (g2SC) phase \cite{SH} or 
gapless CFL (gCFL) phase \cite{gCFL}, depending on the flavor structure of the system. The gapless state resembles 
the unstable Sarma state \cite{Sarma,gABR}. However, under a natural charge neutrality condition, i.e., only 
neutral matter can exist, the gapless phase is indeed a thermal stable state as shown in \cite{SH,gCFL}. 
The existence of thermal stable gapless color superconducting phases was confirmed in Refs.~\cite{gapless-C}
and generalized to finite temperatures in Refs.~\cite{gapless-T}. Recent results based on more careful numerical 
calculations show that the g2SC and gCFL phases can exist at moderate baryon density in the color superconducting 
phase diagram \cite{phase-dia} . 

One of the most important properties of an ordinary superconductor is the Meissner effect, i.e., the 
superconductor expels the magnetic field \cite{Meissner-cond}. In ideal color superconducting 
phases, e.g., in the 2SC and CFL phases, the 
gauge bosons connected with the broken generators obtain masses, which indicates the Meissner screening 
effect \cite{Meissner}. The Meissner effect can be understood using the standard Anderson-Higgs mechanism. 
Unexpectedly, it was found that in the g2SC phase, the Meissner screening masses for five gluons corresponding 
to broken generators of $SU(3)_c$ become imaginary, which indicates a type of chromomagnetic instability in 
the g2SC phase \cite{chromo-ins-g2SC-1,chromo-ins-g2SC-2}. 
The calculations in the gCFL phase show the same type of chromomagnetic instability \cite{chromo-ins-gCFL}.  
Remembering the discovery of superfluidity density instability \cite{Wu-Yip} in the gapless interior-gap 
state \cite{Liu-Wilczek}, it seems that the instability is a inherent property of gapless phases.
(There are several exceptions: 1) It is shown that there is no chromomangetic instability near the 
critical temperature  \cite{no-ins-T} ;  2) It is also found that the gapless phase in strong coupling region 
is free of any instabilities \cite{Strong-gapless}.)

The chromomagnetic instability in the gapless phase still remains as a puzzle. 
By observing that,  the 8-th gluon's chromomagnetic instability  is related to the instability with respect to a 
virtual net momentum of diquark pair, Giannakis and Ren suggested that  a LOFF state might be the true ground 
state \cite{LOFF-Ren-1}. Their further 
calculations show that there is no chromomagnetic instability in a narrow LOFF window when 
the local stability condition is satisfied  \cite{LOFF-Ren-2,GHR-LOFF-Neutral}. 
Latter on, it was found in Ref. \cite{GHM-LOFF-Neutral} that a charge neutral LOFF state 
cannot cure the instability of off-diagonal 4-7th gluons, while a gluon condensate state 
\cite{GHM-gluon} can do the job.  In Ref. \cite{NG-Huang} we further pointed out that, 
when charge neutrality condition is required, there exists another narrow unstable LOFF 
window, not only off-diagonal gluons but the diagonal 8-th gluon cannot avoid the 
magnetic instability. 

In a minimal model of gapless color supercondutor, Hong showed  in Ref. \cite{hong} 
that the mismatch can induce a spontaneous Nambu-Goldstone current generation.
The Nambu-Goldstone current generation state in U(1) case resembles the one-plane
wave LOFF state or diagonal gauge boson's condensate. We  extended the Nambu-Goldstone
current generation picture to the 2SC case  in the nonlinear realization framework in 
Ref. \cite{NG-Huang}.  We show that  five pseudo 
Nambu-Goldstone currents can be spontaneously generated by increasing
the mismatch between the Fermi surfaces of the pairing quarks. The Nambu-Goldstone 
currents generation state covers the gluon phase as well as the 
one-plane wave LOFF state.

This article is organized as follows.  In Sec.~\ref{sec-gNJL}, we describe the framework
of the gauged SU(2) Nambu--Jona-Lasinio (gNJL) model in $\beta$-equilibrium. We review 
chromomagnetic instabilities in Sec.~\ref{sec-Mag-Ins}.  We discuss
 neutral baryon current instability and the LOFF state in Sec.~\ref{sec-bc}.  
Then Sec. ~\ref{sec-NG-current} gives a general Nambu-Goldstone
currents generation description in the non-linearization framework. 
At the end, we give the discussion and summary in Sec. ~\ref{sec-sum}. 

\section{The gauged SU(2) Nambu--Jona-Lasinio model}
\label{sec-gNJL}

We take the gauged form of the extended Nambu--Jona-Lasinio model \cite{Huang-NJL},
the Lagrangian density has the form of
\begin{eqnarray}
\label{lagr}
\mathcal{L} &=& {\bar q}(i\fsl{D}+\hat{\mu}\gamma^0)q  +
   G_S[({\bar q}q)^2 + ({\bar q}i\gamma_5{\bf {\bf \tau}}q)^2 ]  \nonumber \\
 & + & G_D[(i {\bar q}^C  \varepsilon  \epsilon^{b} \gamma_5 q )
   (i {\bar q} \varepsilon \epsilon^{b} \gamma_5 q^C)] ,
\label{lg}
\end{eqnarray}
with $D_\mu \equiv \partial_\mu - ig A_\mu^{a} T^{a}$.  Here $A_\mu^{a}$ 
are gluon fields and  $T^a$ with $a=1,\cdots, 8$ are generators of 
$SU(3)_{\rm c}$ gauge groups.
Please note that we regard all the gauge fields as external fields, which
 are weakly interacting with the system. 
 The property of the color superconducting phase characterized by 
 the diquark gap parameter is determined by the unkown nonperturbative 
 gluon fields, which has been simply replaced by the four-fermion interaction
 in the NJL model.  However, the external gluon fields do not contribute to
 the properties of the system. Therefore, we do not have the contribution to the
 Lagrangian density from gauge field part  ${\cal L}_g$ as introduced in Ref.~\cite{GHM-gluon}.
 (In Sec.~\ref{sec-NG-current}, by using the non-linear realization in the gNJL model,
 we will derive one Nambu-Goldstone currents state, 
 which is equivalent to the so-called gluon-condensate state.)

In the Lagrangian density Eq. (\ref{lg}), $q^C=C {\bar q}^T$, ${\bar q}^C=q^T C$ are charge-conjugate spinors, 
$C=i \gamma^2 \gamma^0$ is the charge conjugation matrix (the superscript $T$ 
denotes the transposition operation). 
The quark field 
$q \equiv q_{i\alpha}$ with $i=u,d$ and $\alpha=r,g,b$ is a flavor 
doublet and color triplet, as well as a four-component Dirac spinor, 
${\bf \tau}=(\tau^1,\tau^2,\tau^3)$ are Pauli matrices in the flavor 
space, where $\tau^2$ is antisymmetric, and $(\varepsilon)^{ik} \equiv \varepsilon^{ik}$,
$(\epsilon^b)^{\alpha \beta} \equiv \epsilon^{\alpha \beta b}$ are totally 
antisymmetric tensors in the flavor and color spaces. 
$\hat{\mu}$ is the matrix of chemical potentials in the color and flavor space.
In $\beta$-equilibrium, the matrix of chemical potentials in the color-flavor 
space ${\hat \mu}$ is given in terms of the quark chemical potential $\mu$, the chemical
potential for the electrical charge $\mu_e$ and the color chemical potential 
$\mu_8$,  
\begin{eqnarray}
\mu_{ij}^{\alpha\beta} = (\mu \delta_{ij} - \mu_e Q_{ij})\delta^{\alpha\beta}
 + \frac{2}{\sqrt{3}} \mu_8 \delta_{ij} (T_8)^{\alpha\beta}.
\end{eqnarray}
$G_S$ and $G_D$ are the quark-antiquark coupling constant 
and the diquark coupling constant, respectively.  In the following, 
we only focus on the color superconducting phase, where
$< {\bar q} q >=0$ and $<{\bar q}\gamma^5 {\bf \tau} q>=0$. 

After bosonization, one obtains the linearized 
version of the model for the 2-flavor superconducting phase,  
\begin{eqnarray}
\label{lagr2}
{\cal L}_{2SC} & =  & {\bar q}(i\fsl{D}+\hat \mu \gamma^0)q 
-\frac{\Delta^{*b}\Delta^{b}}{4G_D}
 \nonumber \\
 &-& \frac{1}{2}\Delta^{*b} (i{\bar q}^C  \varepsilon \epsilon^{b}\gamma_5 q )
  -\frac{1}{2}\Delta^b (i {\bar q}  \varepsilon  \epsilon^{b} \gamma_5 q^C) 
 \label{lagr-2sc}
\end{eqnarray}
with the bosonic fields 
\begin{eqnarray}
 \Delta^b \sim i {\bar q}^C \varepsilon \epsilon^{b}\gamma_5 q, \ \ 
\Delta^{*b} \sim i {\bar q}  \varepsilon  \epsilon^{b} \gamma_5 q^C.
\end{eqnarray}
In the Nambu-Gor'kov space, 
\begin{equation}
  \Psi = \left(\begin{array}{@{}c@{}} q \\ q^C \end{array}\right),
\end{equation}
the inverse of the quark propagator is defined as
\begin{equation}
\left[{\cal S}(P)\right]^{-1} = \left(\begin{array}{cc}
\left[G_0^{+}(P)\right]^{-1} & \Delta^- \\
\Delta^+ & \left[G_0^{-}(P)\right]^{-1}
\end{array}\right),
\label{prop}
\end{equation}
with the off-diagonal elements
\begin{equation}
  \Delta^- \equiv -i\epsilon^b\varepsilon\gamma_5 \Delta, \qquad
  \Delta^+ \equiv -i\epsilon^b\varepsilon\gamma_5 \Delta^*,
\end{equation}
and the free quark propagators $G_0^{\pm}(P)$ taking the form of
\begin{equation}
\left[G_0^{\pm}(P)\right]^{-1} =
 \gamma^0 (p_0 \pm \hat{\mu}) - \vec{\gamma} \cdot \vec{p}.
 \label{freep}
\end{equation}
The 4-momenta are denoted by capital letters, e.g., $P=
(p_0,\vec{p})$. We have assumed the quarks are massless in dense 
quark matter, and the external gluon fields do not contribute to the quark 
self-energy.

The explicit form of the functions $G^{\pm}_I$ and $\Xi^{\pm}_{IJ}$ 
reads
\begin{eqnarray}
\label{quark-propagator}
{\rm G}^{\pm}_1&=&
\frac{(k_0-E_{dg}^{\pm})\gamma^0{\tilde \Lambda}_{k}^+}{(k_0\mp\delta\mu)^2-{E_{\Delta}^{\pm}}^2}
 + \frac{(k_0+E_{dg}^{\mp}) \gamma^0{\tilde \Lambda}_{k}^-}{(k_0\mp\delta\mu)^2-{E_{\Delta}^{\mp}}^2},  
 \nonumber \\
{\rm G}^{\pm}_2&=&
\frac{(k_0-E_{ur}^{\pm})\gamma^0{\tilde \Lambda}_{k}^+}{(k_0\pm\delta\mu)^2-{E_{\Delta}^{\pm}}^2} + \frac{(k_0+E_{ur}^{\mp})\gamma^0{\tilde \Lambda}_{k}^- }{(k_0\pm\delta\mu)^2-{E_{\Delta}^{\mp}}^2}, \nonumber \\
{\rm G}^{\pm}_{3}&=&
\frac{1}{k_0+E_{bu}^{\pm}} \gamma^0{\tilde \Lambda}_{k}^+ + 
\frac{1}{k_0-E_{bu}^{\mp}} \gamma^0{\tilde \Lambda}_{k}^- , \nonumber \\
{\rm G}^{\pm}_{4}&=&
\frac{1}{k_0+E_{bd}^{\pm}} \gamma^0{\tilde \Lambda}_{k}^+ + 
\frac{1}{k_0-E_{bd}^{\mp}} \gamma^0{\tilde \Lambda}_{k}^- , 
\end{eqnarray}
\label{G_I}
with $E^{\pm}_{i\alpha}\equiv E_{k}\pm \mu_{i\alpha}$ and 
\begin{eqnarray}
\Xi^{\pm}_{12} &=& \left(\frac{-i \Delta \gamma^5{\tilde \Lambda}_{k}^+}
{(k_0\pm\delta\mu)^2-{E_{\Delta}^{\pm}}^2} +
\frac{-i \Delta\gamma^5 {\tilde \Lambda}_{k}^- }{(k_0\pm\delta\mu)^2-{E_{\Delta}^{\mp}}^2} \right) ,
\nonumber  \\
\Xi^{\pm}_{21} &=&\left(\frac{ -i \Delta\gamma^5{\tilde \Lambda}_{k}^+ }
{(k_0\mp\delta\mu)^2-{E_{\Delta}^{\pm}}^2} +\frac{ -i \Delta\gamma^5 {\tilde \Lambda}_{k}^- }
{(k_0\mp\delta\mu)^2-{E_{\Delta}^{\mp}}^2} \right), 
\end{eqnarray}
\label{Xi_I}
where 
\begin{eqnarray}
\tilde{\Lambda}^{\pm}_{k} &=& \frac{1}{2}\left(1\pm
\gamma^0\frac{\boldsymbol{\gamma}\cdot\mathbf{k}-m}{E_{k}} \right)\label{t-Lambda-k}
\end{eqnarray}
is an alternative set of energy projectors, and the following notation was used: 
\begin{eqnarray}
E_{k}^{\pm}&\equiv& E_{k} \pm \bar{\mu},\nonumber\\
E_{\Delta,k}^{\pm} &\equiv& \sqrt{(E_{k}^{\pm})^2 +\Delta^2}, \label{g-disp} \\
\bar{\mu} &\equiv&
\frac{\mu_{ur} +\mu_{dg}}{2}
=\frac{\mu_{ug}+\mu_{dr}}{2}
=\mu-\frac{\mu_{e}}{6}+\frac{\mu_{8}}{3}, \nonumber \label{mu-bar}\\
\delta\mu &\equiv&
 \frac{\mu_{dg}-\mu_{ur}}{2}
=\frac{\mu_{dr}-\mu_{ug}}{2}
=\frac{\mu_{e}}{2}. 
\label{delta-mu}
\end{eqnarray}

From the dispersion relation of the quasiparticles Eq. (\ref{g-disp}),  we can 
read that, when $\delta\mu > \Delta$,
there will be excitations of gapless modes in the system. The thermodynamic potential 
corresponding to the solution of the gapless state $\delta\mu > \Delta$ is a local maximum.  
However, under 
certain constraint, e.g., the charge neutrality condition, the gapless 2SC phase can be a 
thermal stable state \cite{SH}.

\section{Chromomagnetic instabilities driven by mismatch}
\label{sec-Mag-Ins}

In this section, we review the chromomagnetic instabilities driven by mismatch in 2SC and g2SC
phases as shown in Ref.~\cite{chromo-ins-g2SC-1,chromo-ins-g2SC-2}. 
The polarization tensor in momentum space has the following general structure:
\begin{equation} 
\Pi^{\mu \nu}_{AB} (P) = 
-\frac{i}{2} \int\frac{d^{4}K}{(2\pi)^{4}}
\mbox{Tr}_{\rm D} \left[ \hat{\Gamma}^\mu_A
{\cal S} (K) \, \hat{\Gamma}^\nu_B {\cal S}(K-P) \right].
\label{PiPscfNG}
\end{equation}
The trace here runs over the Dirac
indices and the vertices 
$\hat{\Gamma}^\mu_A \equiv 
{\rm diag}(g\,\gamma^\mu T_A,-g\,\gamma^\mu T_A^T)$
with $A=1,\ldots,8$.

The Debye masses $m_{D,A}^2$ and the Meissner masses $m_{M,A}^2$ of 
gauge bosons are defined as
\begin{eqnarray}
m_{D,A}^2 &\equiv & - \lim_{p\to 0} {\Pi}_{AA}^{00}(0,p), 
\label{def-Debye}\\
m_{M,A}^2 &\equiv & - \frac{1}{2}\lim_{p\to 0} 
\left(g_{ij}+\frac{p_i p_j}{p^2}\right) {\Pi}_{AA}^{ji}(0,p) .
\label{def-Meissner}
\end{eqnarray}

\subsection{Screening masses of  the gluons $A=1,2,3$}

Gluons $A=1,2,3$ of the unbroken $SU(2)_c$ subgroup 
couple only to the red and green quarks.
The general expression for the polarization tensor 
$\Pi_{AB}^{\mu\nu}(0,p)$ with $A,B=1,2,3$ is diagonal.
After performing the traces over the color, flavor and Nambu-Gorkov indices, 
the expression has the form of
\begin{eqnarray}
\Pi_{11}^{\mu\nu}(P) &=& \frac{g^2T}{4}\sum_n\int \frac{d^3 {\mathbf k}}{(2\pi)^3}
\mbox{Tr}_{\rm D}  \left[
   \gamma^{\mu} G_{1}^+(K) \gamma^{\nu} G_{1}^+(K')  
    + \gamma^{\mu} G_{1}^-(K) \gamma^{\nu}G_{1}^-(K')\right.\nonumber\\
   &+& \left. 
      \gamma^{\mu} G_{2}^+(K) \gamma^{\nu}G_{2}^+(K')  
    + \gamma^{\mu} G_{2}^-(K) \gamma^{\nu}G_{2}^-(K') \right.\nonumber\\
   &+& \left. 
      \gamma^{\mu}\Xi_{12}^-(K) \gamma^{\nu}\Xi_{21}^+(K') 
    + \gamma^{\mu}\Xi_{12}^+(K) \gamma^{\nu}\Xi_{21}^-(K')\right.\nonumber\\
   &+& \left. 
      \gamma^{\mu}\Xi_{21}^-(K) \gamma^{\nu}\Xi_{12}^+ (K')
    + \gamma^{\mu}\Xi_{21}^+(K) \gamma^{\nu}\Xi_{12}^-(K')\right],
\label{Pi11}
\end{eqnarray}  

By making use of the definition in Eq.~(\ref{def-Debye})
and Eq.~(\ref{def-Meissner}).
we arrive at the following result for the threefold degenerate 
Debye mass square:
\begin{eqnarray}
m_{D,1}^2 
&\simeq & \frac{4\alpha_s \bar\mu^2 \delta\mu}
{\pi\sqrt{(\delta\mu)^2-\Delta^2}}\theta(\delta\mu-\Delta),
\label{m_D_1}
\end{eqnarray}
with $\alpha_s\equiv g^2/4\pi$. The Meissner mass square reads
\begin{equation}
m_{M,1}^2 =0. 
\label{m_M_1}
\end{equation}

The Debye screening mass in Eq.~(\ref{m_D_1}) vanishes 
in the gapped phase (i.e., $\Delta/\delta\mu>1$). As in the case of 
the ideal 2SC phase, this reflects the fact that there are no gapless 
quasiparticles charged with respect to the unbroken SU(2)$_c$ gauge 
group. In the gapless 2SC phase, such quasiparticles exist and the 
value of the Debye screening mass is proportional to the density of
states at the corresponding ``effective'' Fermi surfaces.

The Meissner screening mass of the gluons of the unbroken SU(2)$_c$ 
are vanishing in the gapped and gapless 2SC phases.
This is in agreement with the general group-theoretical 
arguments. 

\subsection{Screening masses of diagonal gluon $A=8$}

The 8-th gluon can probe the Cooper-paired red and green quarks, as 
well as the unpaired blue quarks. After the traces over the color, the 
flavor and the Nambu-Gorkov indices are performed, the polarization 
tensor for the 8th gluon can be expressed as 
\begin{subequations}
\begin{eqnarray}
\Pi_{88}^{\mu\nu}(P) &=& \frac{1}{3}{\tilde \Pi}_{88}^{\mu\nu}(P)+  \frac{2}{3}
\Pi_{88,b}^{\mu\nu}(P),  \\ \label{Pi88-2}
{\tilde \Pi}_{88}^{\mu\nu}(P)
&=& \frac{g^2T}{4}\sum_n\int \frac{d^3 {\mathbf k}}{(2\pi)^3}
\mbox{Tr}_{\rm D}  \left[
   \gamma^{\mu} G_{1}^+(K) \gamma^{\nu} G_{1}^+(K')  
    + \gamma^{\mu} G_{1}^-(K) \gamma^{\nu}G_{1}^-(K')\right.\nonumber\\
   &+& \left. 
      \gamma^{\mu} G_{2}^+(K) \gamma^{\nu}G_{2}^+(K')  
    + \gamma^{\mu} G_{2}^-(K) \gamma^{\nu}G_{2}^-(K') \right.\nonumber\\
   &-& \left. 
      \gamma^{\mu}\Xi_{12}^-(K) \gamma^{\nu}\Xi_{21}^+(K') 
    - \gamma^{\mu}\Xi_{12}^+(K) \gamma^{\nu}\Xi_{21}^-(K')\right.\nonumber\\
   &-& \left. 
      \gamma^{\mu}\Xi_{21}^-(K) \gamma^{\nu}\Xi_{12}^+ (K')
    - \gamma^{\mu}\Xi_{21}^+(K) \gamma^{\nu}\Xi_{12}^-(K')\right],
\label{Pi88} \\ 
\Pi_{88,b}^{\mu\nu}(P) &=&  \frac{g^2T}{4}\sum_n\int \frac{d^3 {\mathbf k}}{(2\pi)^3}
\mbox{Tr}_{\rm D}  \left[ \gamma^{\mu} G_{3}^+(K) \gamma^{\nu}
G_{3}^+(K')  + \gamma^{\mu} G_{3}^-(K) \gamma^{\nu}G_{3}^-(K')
\right.\nonumber\\
   &+& \left. \gamma^{\mu} G_{4}^+(K) 
\gamma^{\nu}G_{4}^+(K')  + \gamma^{\mu} G_{4}^-(K) \gamma^{\nu}G_{4}^-(K')
\right].
\label{Pi88b}
\end{eqnarray}
\end{subequations}

The expressions for the Debye 
screening mass reads
\begin{eqnarray}
m_{D,8}^2 &=& \frac{4\alpha_s\bar\mu^2}{\pi} ,
\label{m_D_8} 
\end{eqnarray}
and the Meissner screening mass takes the form of  
\begin{eqnarray}
m_{M,88}^2 &=& \frac{4\alpha_s\bar\mu^2}{9\pi} 
\left(1-\frac{\delta\mu~\theta(\delta\mu-\Delta)}
{\sqrt{(\delta\mu)^2-\Delta^2}} \right).
\label{m_M_88} 
\end{eqnarray}

As is easy to see from Eq.~(\ref{m_M_88}), the 
Meissner screening mass squared of 8-th gluon is 
{\em negative} when $0<\Delta/\delta\mu<1$, this 
indicates a magnetic plasma instability in the gapless 
2SC phase. 

\subsection{Screening masses of the gluons with $A=4,5,6,7$}

After performing the traces over the color, the flavor and the Nambu-Gorkov indices,
the diagonal components of the polarization tensor $\Pi_{AB}^{\mu\nu}(P)$ with $A=B=4,5,6,7$
have the form
\begin{eqnarray}
\Pi_{44}^{\mu\nu}(P) & = & 
\frac{g^2 T}{8} \int \frac{d^3 {\mathbf k}}{(2\pi)^3}{\rm Tr}_{\rm D}
\big[ \gamma^{\mu} G_{3}^+(K) \gamma^{\nu}G_{1}^+(K') 
+ \gamma^{\mu} G_{1}^+(K) \gamma^{\nu} G_{3}^+(K') \nonumber \\
& + & \gamma^{\mu} G_{4}^+(K) \gamma^{\nu}G_{2}^+(K')
+ \gamma^{\mu} G_{2}^+(K) \gamma^{\nu}G_{4}^+(K') \nonumber \\
& + & \gamma^{\mu} G_{3}^-(K) \gamma^{\nu}G_{1}^-(K') 
+ \gamma^{\mu} G_{1}^-(K) \gamma^{\nu}G_{3}^-(K')\nonumber \\
& + & \gamma^{\mu} G_{4}^-(K) \gamma^{\nu}G_{2}^-(K') 
+ \gamma^{\mu} G_{2}^-(K) \gamma^{\nu}G_{4}^-(K') \big].
\end{eqnarray}
Note that $\Pi_{44}^{\mu\nu}(P)=\Pi_{55}^{\mu\nu}(P)
=\Pi_{66}^{\mu\nu}(P)=\Pi_{77}^{\mu\nu}(P)$.
Apart from the diagonal elements, there are also nonzero 
off-diagonal elements,
\begin{eqnarray}
\Pi_{45}^{\mu\nu}(P) = - \Pi_{54}^{\mu\nu} (P)
= \Pi_{67}^{\mu\nu}(P) = - \Pi_{76}^{\mu\nu}(P)
= i {\hat \Pi}^{\mu\nu}(P), 
\label{odd-diag}
\end{eqnarray}
with 
\begin{eqnarray}
{\hat \Pi}^{\mu\nu}(P) &=& 
\frac{g^2 T}{8} \int \frac{d^3 {\mathbf k}}{(2\pi)^3}{\rm Tr}_{\rm D}
\big[ \gamma^{\mu} G_{3}^+(K) \gamma^{\nu}G_{1}^+(K') 
- \gamma^{\mu} G_{1}^+(K) \gamma^{\nu} G_{3}^+(K') \nonumber \\
&+& \gamma^{\mu} G_{4}^+(K) \gamma^{\nu}G_{2}^+(K')
- \gamma^{\mu} G_{2}^+(K) \gamma^{\nu}G_{4}^+(K') \nonumber \\
& - & \gamma^{\mu} G_{3}^-(K) \gamma^{\nu}G_{1}^-(K') 
+ \gamma^{\mu} G_{1}^-(K) \gamma^{\nu}G_{3}^-(K')\nonumber \\
& - & \gamma^{\mu} G_{4}^-(K) \gamma^{\nu}G_{2}^-(K') 
+ \gamma^{\mu} G_{2}^-(K) \gamma^{\nu}G_{4}^-(K') \big].
\end{eqnarray}
The physical gluon fields in the 2SC/g2SC phase
are the following linear combinations:
$\tilde{A}_{4,5}^{\mu}=(A_{4}^{\mu}\pm i A_{5}^{\mu})/\sqrt{2}$
and  $\tilde{A}_{6,7}^{\mu}=(A_{6}^{\mu}\pm i A_{7}^{\mu})/\sqrt{2}$. 
These new fields, 
$\tilde{A}_{4,6}^{\mu}$ and $\tilde{A}_{7,5}^{\mu}$
describe two pairs of massive vector particles with well defined 
electomagnetic charges, $\tilde{Q}=\pm 1$. 
The components of the polarization tensor in the new
basis read
\begin{subequations}
\begin{eqnarray}
\tilde{\Pi}_{44}^{\mu\nu}(P) & = & \tilde{\Pi}_{66}^{\mu\nu}(P) = \Pi_{44}^{\mu\nu}(P) + 
{\hat \Pi}^{\mu\nu}(P)  \nonumber \\
& = & 
\frac{g^2 T}{4} \int \frac{d^3 {\mathbf k}}{(2\pi)^3}{\rm Tr}_{\rm D}
 \big[ 
\gamma^{\mu} G_{3}^+(K)
 \gamma^{\nu}G_{1}^+(K')  + \gamma^{\mu} G_{1}^-(K) 
\gamma^{\nu}G_{3}^-(K') \nonumber \\
& & + \gamma^{\mu} G_{4}^+(K) \gamma^{\nu}G_{2}^+(K') + 
\gamma^{\mu} G_{2}^-(K) \gamma^{\nu}G_{4}^-(K') \big], 
\end{eqnarray}
and 
\begin{eqnarray}
\tilde{\Pi}_{55}^{\mu\nu}(P) & = & \tilde{\Pi}_{77}^{\mu\nu}(P) = \Pi_{44}^{\mu\nu} (P)
- {\hat \Pi}^{\mu\nu}(P) \nonumber \\
& = & 
\frac{g^2 T}{4} \int \frac{d^3 {\mathbf k}}{(2\pi)^3}{\rm Tr}_{\rm D}
 \big[ 
\gamma^{\mu} G_{1}^+(K)\gamma^{\nu}G_{3}^+(K')  + \gamma^{\mu} 
G_{3}^-(K) \gamma^{\nu}G_{1}^-(K') \nonumber \\
& & + \gamma^{\mu} G_{2}^+(K) \gamma^{\nu}G_{4}^+(K') + 
\gamma^{\mu} G_{4}^-(K) \gamma^{\nu}G_{2}^-(K') \big].
\end{eqnarray}
\end{subequations}

In the static limit, all four eigenvalues of the polarization tensor 
are degenerate. By making use of the definition in Eq.~(\ref{def-Debye}), 
we derive the following result for the corresponding Debye masses:
\begin{equation}
m_{D,4}^2 = \frac{4\alpha_s\bar\mu^2}{\pi} \left[
\frac{\Delta^2+2\delta\mu^2}{2\Delta^2} 
-\frac{\delta\mu\sqrt{\delta\mu^2-\Delta^2}}{\Delta^2}
\theta(\delta\mu-\Delta)\right] \!\! .
\label{m_D_4}
\end{equation}
Here we assumed that $\mu_8$ is vanishing which is a good 
approximation in neutral two-flavor quark matter.

The fourfold degenerate Meissner screening mass of the gluons 
with $A=4,5,6,7$ reads
\begin{equation}
m_{M,4}^2 = \frac{4\alpha_s\bar\mu^2}{3\pi} \left[
\frac{\Delta^2-2\delta\mu^2}{2\Delta^2} 
+\frac{\delta\mu\sqrt{\delta\mu^2-\Delta^2}}{\Delta^2}
\theta(\delta\mu-\Delta)\right] \!\! .
\label{m_M_4}
\end{equation}
Both results in Eqs.~(\ref{m_D_4}) and (\ref{m_M_4}) interpolate between 
the known results in the normal phase (i.e., $\Delta/\delta\mu=0$) and 
in the ideal 2SC phase (i.e., $\Delta/\delta\mu=\infty$). 
The instability of off-diagonal gluons appears in the whole region of 
gapless 2SC phases (with $0<\Delta/\delta\mu<1$) and even in some 
gapped 2SC phases (with $1<\Delta/\delta\mu<\sqrt{2}$). 

It is noticed that the Meissner screening mass square for the off-diagonal
gluons decreases monotonously to zero  when the mismatch increases
from zero to $ \delta\mu= \Delta/\sqrt{2}$, then goes to negative value
with further increase of the mismatch in the gapped 2SC phase.  
However, the behavior of diagonal 8-th gluon's Meissner mass square
is quite different. It keeps as a constant in the gapped 2SC phase. 
In the gapless 2SC phase, all these five gluons' Meissner mass square
are negative. 

\section{Color neutral baryon current instability}
\label{sec-bc}

It is not understood why gapless color superconducting phases exhibit chromomagnetic instability. It sounds 
quite strange especially  in the g2SC phase, where it is the electrical neutrality not the color neutrality playing
the essential role. It is a puzzle why the gluons can feel the instability by requiring the electrical neutrality on 
the system. In order to understand what is really going `wrong' with the homogeneous g2SC phase, we want 
to know whether there exists other instabilities except the chromomagnetic instability. For that purpose, we 
probe the g2SC phase using different external sources,  e.g., scalar and vector diquarks, mesons, vector 
current, and so on.  Here we report the most interesting result regarding the response of the 
g2SC phase to an external vector current $V^{\mu}={\bar \psi} \gamma^{\mu} \psi$, the time-component and 
spatial-components of this current correspond to the baryon number density and baryon current, respectively.

From the linear response theory, the induced current and the external vector current is related by the 
response function
 $\Pi^{\mu\nu}_{V}(P)$, 
 \begin{equation} \label{PiVNG}
\Pi^{\mu \nu}_{V} (P) = \frac{1}{2} \, \frac{T}{V}
\sum_K {\rm Tr} \left[ \hat{\Gamma}^\mu_V
{\cal S} (K) \hat{\Gamma}^\nu_V {\cal S}(K-P) \right] .
\end{equation}
The trace here runs over the Nambu-Gorkov, flavor, color and Dirac
indices.  The explicit form of vertices is $\hat{\Gamma}_V^\mu \equiv  {\rm diag}( \gamma^\mu, -\gamma^\mu)$. 

The explicit expression of the vector current response function takes the form of 
\begin{eqnarray}
\Pi_{V}^{\mu\nu}(P) &=& {\tilde \Pi}_{V}^{\mu\nu}(P)+ \Pi_{V,b}^{\mu\nu}(P),  
\label{PiV-2}
\end{eqnarray}
with 
\begin{eqnarray}
{\tilde \Pi}_{V}^{\mu\nu}(P)
&=& \frac{T}{2}\sum_n\int \frac{d^3 {\mathbf k}}{(2\pi)^3}
\mbox{Tr}_{\rm D}  \left[ \right. \nonumber \\
   & & \left.  \gamma^{\mu} G_{1}^+(K) \gamma^{\nu} G_{1}^+(K')  
    + \gamma^{\mu} G_{1}^-(K) \gamma^{\nu}G_{1}^-(K')\right.\nonumber\\
   &+& \left. 
      \gamma^{\mu} G_{2}^+(K) \gamma^{\nu}G_{2}^+(K')  
    + \gamma^{\mu} G_{2}^-(K) \gamma^{\nu}G_{2}^-(K') \right.\nonumber\\
   &-& \left. 
      \gamma^{\mu}\Xi_{12}^-(K) \gamma^{\nu}\Xi_{21}^+(K') 
    - \gamma^{\mu}\Xi_{12}^+(K) \gamma^{\nu}\Xi_{21}^-(K')\right.\nonumber\\
   &-& \left. 
      \gamma^{\mu}\Xi_{21}^-(K) \gamma^{\nu}\Xi_{12}^+ (K')
    - \gamma^{\mu}\Xi_{21}^+(K) \gamma^{\nu}\Xi_{12}^-(K')\right], \\
\Pi_{V,b}^{\mu\nu}(P) &=&  \frac{T}{2}\sum_n\int \frac{d^3 {\mathbf k}}{(2\pi)^3}
\mbox{Tr}_{\rm D}  \left[ \right. \nonumber\\
 & & \left. \gamma^{\mu} G_{3}^+(K) \gamma^{\nu}
G_{3}^+(K')  + \gamma^{\mu} G_{3}^-(K) \gamma^{\nu}G_{3}^-(K')
\right.\nonumber\\
   &+& \left. \gamma^{\mu} G_{4}^+(K) 
\gamma^{\nu}G_{4}^+(K')  + \gamma^{\mu} G_{4}^-(K) \gamma^{\nu}G_{4}^-(K')
\right], 
\end{eqnarray}
here the trace is over the Dirac space.

Comparing the explicit expression of $\Pi_{V}^{\mu\nu}(P)$ with that of the 8-th gluon's 
self-energy $\Pi_{88}^{\mu\nu}(P)$, i.e., Eq. (\ref{Pi88}), it can be 
clearly seen that,  $\Pi_{V}^{\mu\nu}(P)$  and $\Pi_{88}^{\mu\nu}(P)$ almost share the same 
expression, except the coefficients.  This can be easily understood, because the color charge 
and color current carried by the 8-th gluon is proportional to the baryon number and baryon 
current, respectively. In the static long-wavelength ($p_0=0$ and $\vec{p} \to 0$ ) limit, the 
 time-component and spatial component of $\Pi_{V}^{\mu\nu}(P)$
give the baryon number susceptibility $\xi_n$ and baryon current susceptibility $\xi_c$, respectively,
\begin{eqnarray}
\xi_n  &\equiv & - \lim_{\vec{p}\to 0} \tilde{\Pi}_{V}^{00}(0,\vec{p}) \propto m_{8,D}^2, 
\label{def-xin}\\
\xi_c &\equiv & - \frac{1}{2}\lim_{\vec{p}\to 0} \left(g_{ij}+\frac{p_i p_j}{p^2}\right) {\Pi}_{V}^{ij}(0,\vec{p}) \propto m_{8,M}^2.
\label{def-xic}
\end{eqnarray}

In the g2SC phase, $m_{8,M}^2$ as well as $\xi_c$ become negative. This means that, except the 
chromomagnetic instability corresponding to broken generators of $SU(3)_c$, and the instability
of a net momentum for diquark pair, the g2SC phase is 
also unstable with respect to an external color neutral baryon current $\bar{\psi}\vec{\gamma}\psi$.

The 8-th gluon's magnetic instability, the diquark momentum instability and the color neutral
baryon current in the g2SC phase can be understood in one common physical picture.  
The g2SC phase exhibits a paramagnetic response to an external baryon current. 
Naturally, the color current carried by the 8-th gluon, which differs from the baryon current by a 
color charge, also experiences the instability in the g2SC phase.  The paramagnetic instability 
of the baryon current indicates that the quark can spontaneously obtain a momentum, because 
diquark carries twice of the quark momentum, it is not hard to understand why the g2SC phase 
is also unstable with respect to the response of a net diquark momentum. 

It is noticed that, the instability of $\bar{\psi}\vec{\gamma}\psi$ will be induced by mismatch in all 
the asymmetric Fermi pairing systems, including superfluid systems,  where 
$\bar{\psi}\vec{\gamma}\psi$ can be interpreted as particle current. 

\subsection{Spontaneous baryon current generation}
\label{sec-UsLOFF}

The paramagnetic response to an external vector current naturally suggests that a vector current can be 
spontaneously generated in the system. The generated vector current behaves as a vector potential, 
which modifies the quark self-energy 
with a spatial vector condensate $\vec{\gamma} \cdot \vec{\Sigma}_V$, and breaks the rotational symmetry 
of the system. It can also be understood that the quasiparticles
in the gapless phase spontaneously obtain a superfluid velocity, and the ground state is in an anisotropic state. The quark propagator $G_0^{\pm}(P)$ in Eq. (\ref{freep}) is modified as 
\begin{equation}
\left[G_{0,V}^{\pm}(P)\right]^{-1} =
 \gamma^0 (p_0 \pm \hat{\mu}) - \vec{\gamma} \cdot \vec{p} \mp \vec{\gamma} \cdot \vec{\Sigma}_V,
 \label{freep-m}
\end{equation}
with a subscript $V$ indicating the modified quark propagator. 
Correspondingly, the  inverse of the quark propagator $[{\cal S}(P)]^{-1}$ in Eq. (\ref{prop}) is modified as
\begin{equation}
\left[{\cal S}_V(P)\right]^{-1} = \left(\begin{array}{cc}
\left[G_{0,V}^{+}(P)\right]^{-1} & \Delta^- \\
\Delta^+ & \left[G_{0,V}^{-}(P)\right]^{-1}
\end{array}\right).
\label{prop-m}
\end{equation}
It is noticed that the expression of the modified inverse quark propagator $[S_V(P)]^{-1}$ takes the same form as the inverse quark propagator in the one-plane wave LOFF state shown in Ref. \cite{LOFF-Ren-2}. The net momentum $\vec{q}$ of the diquark pair in 
the LOFF state \cite{LOFF-Ren-2} is replaced here by a spatial vector condensate $\vec{\Sigma}_V$.  
The spatial vector condensate $\vec{\gamma} \cdot \vec{\Sigma}_V$ breaks rotational symmetry of the system. 
This means that the Fermi surfaces of the pairing quarks are not spherical any more.

It has to be pointed out, the baryon current offers one Doppler-shift superfluid velocity for the 
quarks. A spontaneously generated Nambu-Goldstone current in the minimal gapless model \cite{hong} 
or a condensate of 8-th gluon's spatial component can do the same job. All these states mimic the 
one-plane wave LOFF state. In the following, we just call all these states as the single-plane wave LOFF state. 

 In order to determine the 
deformed structure of the Fermi surfaces,  one should self-consistently minimize the free energy $\Gamma(\Sigma_V, \Delta, \mu,\mu_e,\mu_8)$. The explicit form of the free energy can be evaluated 
 directly using the standard method, in the framework of Nambu--Jona-Lasinio model \cite{neutral_huang,SH},  
 it takes the form of
\begin{eqnarray}
\Gamma= - \frac{T}{2} \sum_n\int\frac{d^3\vec{p}}{(2\pi)^3} {\rm Tr} \ln ( [{\cal S}_V(P)]^{-1}) + \frac{\Delta^2}{4 G_D},
\end{eqnarray}
where $T$ is the temperature, and $G_D$ is the coupling constant in the diquark channel. 

When there is no charge neutrality condition, the ground state is determined
by the thermal stability condition, i.e., the local stability condition. The ground state is in the 
2SC phase when $\delta\mu<0.706 \Delta_0$ with $\Delta \simeq \Delta_0$, in the LOFF phase when 
$0.706 \Delta_0 < \delta\mu <0.754 \Delta_0$ correspondingly $ 0<\Delta/\Delta_0<0.242$, 
and then in the normal phase with $\Delta=0$ when the mismatch is larger than $0.754 \Delta_0$. 
Here $\Delta, \Delta_0$ indicate the diquark gap in the case of $\delta\mu \neq 0$ and $\delta\mu=0$,
respectively.

\subsection{Unstable neutral LOFF window}

Now come to the charge neutral LOFF state, and investigate whether the LOFF
state can resolve all the magnetic instabilities.
 
When charge neutrality condition is required, the ground state of charge neutral 
quark matter should be determined by solving the gap equations  as well as the 
charge neutrality condition, i.e., 
\begin{eqnarray}
\frac{\partial \Gamma}{\partial \Sigma_V} =0, \ \, \frac{\partial \Gamma}{\partial \Delta} =0, \ \
\frac{\partial \Gamma}{\partial \mu_e} =0, \ \, \frac{\partial \Gamma}{\partial \mu_8} =0.
\end{eqnarray}
By changing $\Delta_0$ or coupling strength $G_D$, the solution of the charge neutral 
LOFF state can stay everywhere in the full LOFF window, including the window not 
protected by the local stability condition, as shown explicitly in Ref. \cite{GHM-LOFF-Neutral}.  

From the lesson of charge neutral g2SC phase, we learn that
even though the neutral state is a thermal stable state, i.e., the thermodynamic potential is
a global minimum along the neutrality line, it cannot guarantee the
dynamical stability of the system. The stability of the neutral system
should be further determined by the dynamical stability condition, i.e., 
the positivity of the Meissner mass square. 

The polarization tensor for the gluons with color $A=4,5,6,7,8$ should be evaluated using 
the modified quark propagator ${\cal S}_V$ in Eq. (\ref{prop-m}), i.e.,
\begin{equation} 
\Pi^{\mu \nu}_{AB} (P) = \frac{1}{2} \, \frac{T}{V} \sum_K
\mbox{Tr} \left[ \hat{\Gamma}^\mu_A
{\cal S}_V (K) \, \hat{\Gamma}^\nu_B {\cal S}_V (K-P) \right],
\label{PiAB}
\end{equation}
with $A,B=4,5,6,7,8$ and the explicit form of the vertices $\hat{\Gamma}^\mu_A$ has the form
$\hat{\Gamma}_A^\mu
\equiv {\rm diag}(g\,\gamma^\mu T_A,-g\,\gamma^\mu T_A^T) $.
In the LOFF state, the Meissner tensor can be decomposed into transverse and longitudinal
component. The transverse and longitudinal Meissner mass square for the off-diagonal 
4-7 gluons and the diagonal 8-th gluon have been performed explicitly in the one-plane wave
LOFF state in Ref. \cite{LOFF-Ren-2}. 

According to the dynamical stability condition,
i.e., the positivity of the transverse as well as longitudinal Meissner mass 
square, we can devide the LOFF state into three LOFF windows \cite{Hai-cang}:

1) The stable LOFF (S-LOFF) window in the region of  
$0<\Delta/\Delta_0<0.39$, which is free of any magnetic instability.
Please note that this S-LOFF window is a little bit wider than the window 
$0<\Delta/\Delta_0<0.242$ protected by the local stability condition.

2) The stable window for diagonal gluon characterized by Ds-LOFF window
in the region of  $0.39<\Delta/\Delta_0<0.83$, which is free of the diagonal 8-th gluon's 
magnetic instability but not free of the off-diagonal gluons' magnetic instability;

3) The unstable LOFF (Us-LOFF) window in the region of 
$0.83<\Delta/\Delta_0< r_c$, with $r_c \, \equiv \, \Delta(\delta\mu=\Delta)/\Delta_0 \simeq 1$.
In this Us-LOFF window,  all the magnetic instabilities exist. Please note that, it is
the longitudinal Meissner mass square for the 8-th gluon is negative in this Us-LOFF
window, the transverse Meissner mass square of 8-th gluon is always zero in the full
LOFF window, which is guaranteed by the momentum equation.  

Us-LOFF is a very interesting window, it indicates that the LOFF state even 
cannot cure the 8-th gluon's magnetic instability.
In the charge neutral 2-flavor system, it seems that the diagonal gluon's magnetic
instability cannot be cured in the gluon phase, because there is no direct relation
between the diagonal gluon's instability and the off-diagonal gluons'  instability.
(Of course, it has to be carefully checked, whether 
all the instabilities in this Us-LOFF window can be cured by 
off-diagonal gluons' condensate in the charge neutral 2-flavor system.)
It is also noticed that in this Us-LOFF window, the mismatch is close to the diquark gap, i.e.,
$\delta\mu \simeq \Delta$. Therefore it is interesting to check whether this Us-LOFF window
can be stabilized by a spin-1 condensate \cite{spin-1} as proposed in Ref. \cite{hong}. 

In the charge neutral 2SC phase, though it is unlikely, we might have a lucky chance
to cure the diagonal instability by the condensation of off-diagonal gluons.
It is expected that this instability will show up in some constrained Abelian asymmetric 
superfluidity system, e.g., in the fixed number density case \cite{He-Jin-Zhuang}.
It will be a new challenge for us to really solve this problem. 

\section{Spontaneous Nambu-Goldstone currents generation}
\label{sec-NG-current}

We have seen that, except chromomagnetic instability corresponding to broken 
generators of $SU(3)_c$, the g2SC phase is also unstable with respect to
the external neutral baryon current. It is noticed that all the instabilities are 
induced by increasing the mismatch between the Fermi surfaces of the 
Cooper pairing. In order to understand the instability driven by mismatch, 
in the following, we give some general analysis.

A superconductor will be eventually destroyed and goes to the normal Fermi liquid state,
so one natural question is: how an ideal BCS superconductor will be destroyed by increasing 
mismatch? To answer how a superconductor will be destroyed, one has to 
firstly understand what is a superconductor. The superconducting phase 
is characterized by the order parameter  $\Delta(x)$, which is a complex scalar field 
and has the form of e.g., for electrical superconductor, 
$\Delta (x) = |\Delta| e^{i\varphi (x)}$, with $|\Delta| $ the amplitude and $\varphi$ the phase
of the gap order parameter or the pseudo Nambu-Goldstone boson. 

1) The superconducting phase is charaterized by the nonzero vacuum expectation value, i.e., 
$<\Delta>\neq 0$, which means the amplitude of the gap is finite, and the phase coherence 
is also established.
 
2) If the amplitude is still finite, while the phase coherence is lost, this phase
is in a phase decoherent  pseudogap state characterized by $|\Delta|\neq 0$, 
but  $<\Delta> =0$ because of $ <e^{i\varphi(x)}> = 0$.

3) The normal state is characterized by $|\Delta|=0$. 

There are two ways to destroy a superconductor.  One way is by driving the 
amplitude of the order parameter to zero. This way is BCS-like, because it mimics the behavior of a 
conventional superconductor at finite temperature, the gap amplitude monotonously 
drops to zero with the increase of temperature;
Another way is non-BCS like, but Berezinskii-Kosterlitz-Thouless (BKT)-like \cite{BKT},  
even if the amplitude of the order parameter is large and finite, superconductivity will be 
lost with the destruction of phase coherence, e.g.  the phase transition from the $d-$wave
superconductor to the pseudogap state in high temperature superconductors 
\cite{Emery-Kivelson}. 

Stimulating by the role of the phase fluctuation in the unconventional superconducting
phase in condensed matter, we follow Ref. \cite{NonLinear} to formulate the 2SC phase 
in the nonlinear realization framework in order to naturally take into account the contribution 
from the phase fluctuation or pseudo Nambu-Goldstone current. 

In the 2SC phase, the color symmetry $G=SU(3)_c$  breaks to $H=SU(2)_c$.
The generators of the residual $SU(2)_c$ symmetry H are 
$\{S^a=T^a\}$ with $a=1,2,3$ and the broken generators $\{X^b=T^{b+3}\}$ 
with $b=1, \cdots, 5$. More precisely, the last broken 
generator is a combination of $T_8$ and the generator ${\bf 1}$ 
of the global $U(1)$ symmetry of baryon number conservation, 
$B \equiv ({\bf 1} + \sqrt{3} T_8)/3$ of generators of
the global $U(1)_B$ and local $SU(3)_c$ symmetry.

The coset space $G/H$ is parameterized by the group elements
\begin{equation} \label{phase}
{\cal V}(x) \equiv \exp \left[ i \left( \sum_{a=4}^7 \varphi_a(x) T_a
+ \frac{1}{\sqrt{3}}\, \varphi_8(x) B \right) \right]\,\,,
\end{equation}
here $\varphi_a (a=4,\cdots,7)$ and $\varphi_8$ are five Nambu-Goldstone 
diquarks, and we have neglected the singular phase, which should include the
information of the topological defects \cite{FT,Topo}.
Operator ${\cal V}$ is unitary, ${\cal V}^{-1} = {\cal V}^\dagger$.

Introducing a new quark field $\chi$, which is connected with the original 
quark field $q$ in Eq. (\ref{lagr-2sc}) in a nonlinear transformation form,
\begin{equation} \label{chi}
q = {\cal V}\, \chi
\,\,\,\, , \,\,\,\,\,
\bar{q} = \bar{\chi}\, {\cal V}^\dagger\,\, ,
\end{equation}
and the charge-conjugate fields transform as
\begin{equation}
q_{C} = {\cal V}^* \, \chi_{C}
\,\,\,\, , \,\,\,\,\,
\bar{q}_{C} = \bar{\chi}_{C} \, {\cal V}^T\,\, .
\end{equation}
In high-$T_c$ superconductor, this technique is called
charge-spin separation, see Ref. \cite{FT}. 
The advantage of transforming the quark fields is
that this preserves the simple structure of the terms coupling
the quark fields to the diquark sources,
\begin{equation}
\bar{q}_{C}\, \Delta^+ \, q
\equiv \bar{\chi}_{C}\, \Phi^+ \, \chi
\,\,\,\, , \,\,\,\,\,
\bar{q}\, \Delta^- \, q_{C}
\equiv \bar{\chi} \, \Phi^- \, \chi_{C} \,\, .
\end{equation}
In mean-field approximation, the diquark source
terms are proportional to
\begin{equation} \label{mfa2}
\Phi^+ 
\sim \langle \, \chi_{C} \, \bar{\chi}\, \rangle 
\,\,\,\, , \,\,\,\,\,
\Phi^- 
\sim \langle \, \chi \, \bar{\chi}_{C} \, \rangle\,\, .
\end{equation}

Introducing the new Nambu-Gor'kov spinors
\begin{equation}
X \equiv \left( \begin{array}{c} 
                    \chi \\
                    \chi_{C} 
                   \end{array}
            \right) \,\,\, , \,\,\,\,
\bar{X} \equiv ( \bar{\chi} \, , \, \bar{\chi}_{C} ),
\end{equation}
the nonlinear realization of the original Lagrangian density
Eq.(\ref{lagr-2sc}) takes the form of
\begin{equation}
{\cal L}_{nl} \equiv
 \bar{X} \,  {\cal S}_{nl}^{-1} \, X 
- \frac{\Phi^+\Phi^-}{4 G_D} \,\, ,
\label{lagr-nl}
\end{equation}
where 
\begin{equation}
{\cal S}_{nl}^{-1} \equiv 
\left( \begin{array}{cc}
            [G^+_{0,nl}]^{-1} & \Phi^- \\
             \Phi^+ & [G^-_{0,nl}]^{-1}
       \end{array} \right)\,\, .
\end{equation}
Here the explicit form of the free propagator for the new quark field is
\begin{eqnarray}
[G^+_{0,nl}]^{-1} & = & i\, \fsl{D} + {\hat \mu} \, \gamma_0 + \gamma_\mu \, V^\mu, 
\end{eqnarray}
and
\begin{eqnarray}
[G^-_{0,nl}]^{-1} & = & i\, \fsl{D}^T - {\hat \mu} \, \gamma_0 + \gamma_{\mu} \, V_C^\mu .
\end{eqnarray}
Comparing with the free propagator in the original Lagrangian density, 
the free propagator in the non-linear realization framework naturally takes
into account the contribution from the Nambu-Goldstone currents or 
phase fluctuations, i.e.,
\begin{eqnarray}
V^\mu & \equiv & {\cal V}^\dagger \, \left( i \, \partial^\mu \right) \, {\cal V}, \nonumber \\
V^\mu_C & \equiv & {\cal V}^T \, \left( i \, \partial^\mu \right) \, {\cal V}^*,
\end{eqnarray}
which is the $N_c N_f \times N_c N_f$-dimensional
Maurer-Cartan one-form introduced in Ref. \cite{NonLinear}.
The linear order of the Nambu-Goldstone currents 
$V^\mu$ and $V_C^\mu $ has the explicit form of
\begin{eqnarray}
V^\mu & \simeq &  - \sum_{a=4}^7 
\left( \partial^\mu  \varphi_a \right)\, T_a - 
\frac{1}{\sqrt{3}}\, \left(\partial^\mu  \varphi_8\right)\, B\,\, , \\
V_C^\mu & \simeq &   \sum_{a=4}^7 
\left( \partial^\mu \varphi_a \right) \, T_a^T + 
\frac{1}{\sqrt{3}}\, \left( \partial^\mu  \varphi_8\right) \, B^T\,\, .
\end{eqnarray}

The Lagrangian density Eq. (\ref{lagr-nl}) for the new quark fields looks like an 
extension of the theory in Ref. \cite{FT} for high-$T_c$ superconductor to Non-Abelian
system, except that here we neglected the singular phase
contribution from the topologic defects. 
The advantage of the non-linear realization framework Eq. (\ref{lagr-nl}) is
that it can naturally take into account the contribution from the phase fluctuations 
or Nambu-Goldstone currents. 

The task left is to correctly solve the ground state by 
considering the phase fluctuations. 
The free energy $\Gamma(V_{\mu},\Delta, \mu,\mu_8, \mu_e)$ can be evaluated 
 directly and it  takes the form of
\begin{eqnarray}
\Gamma= - \frac{T}{2} \sum_n\int\frac{d^3\vec{p}}{(2\pi)^3} {\rm Tr} \ln ( [{\cal S}_{nl}(P)]^{-1}) + \frac{\Phi^2}{4 G_D}.
\label{free-energy-NG}
\end{eqnarray}
To evaluate the ground state of $\Gamma(V_{\mu},\Delta, \mu,\mu_8, \mu_e)$ as a function of mismatch is tedious 
and still under progress. 
In the following we just give a brief discussion on the Nambu-Goldstone current generation 
state \cite{hong}, one-plane wave LOFF state \cite{LOFF-Ren-1,LOFF-Ren-2}, 
as well as the gluon phase \cite{GHM-gluon}.

If we expand the thermodynamic potential $\Gamma(V_{\mu},\Delta, \mu,\mu_8, \mu_e)$ of the non-linear realization form
in terms of the Nambu-Goldstone currents, we will naturally have the Nambu-Goldstone
currents generation in the system with the increase of mismatch, i.e.,
$<\sum_{a=4}^7 {\vec \triangledown}  \varphi_a>\neq 0$ and/or $<{\vec \triangledown}  \varphi_8 > \neq 0$
at large $\delta\mu$. This is an extended version of the Nambu-Goldstone current generation
state proposed in a minimal gapless model in Ref. \cite{hong,Huang-PKU}. 
From Eq. (\ref{lagr-nl}), we can see that ${\vec \triangledown}  \varphi_8$ contributes 
to the baryon current. $<{\vec \triangledown}  \varphi_8 > \neq 0$ indicates a
baryon current generation or 8-th gluon condensate in the system, it 
is just the one-plane wave LOFF state. This has been discussed in 
Sec. \ref{sec-UsLOFF}. The other four
Nambu-Goldstone currents generation 
$<\sum_{a=4}^7 {\vec \triangledown}  \varphi_a>\neq 0$ indicates
other color current generation in the system, and
is equivalent to the gluon phase described in Ref. \cite{GHM-gluon}. 

We do not argue whether the system will exprience 
a gluon condensate phase or Nambu-Goldstone currents generation
state. We simply think they are equivalent. In fact, the gauge fields and
the Nambu-Goldstone currents share a gauge covariant form 
as shown in the free propogator. However, we prefer to using 
Nambu-Goldstone currents generation than the gluon condensate 
in the gNJL model. As mentioned in Sec. \ref{sec-gNJL}, 
in the gNJL model, all the information from unkown nonperturbative 
gluons are hidden in the diquark gap parameter $\Delta$.  The gauge fields in 
the Lagrangian density are just external fields, they only play the 
role of probing the system, but do not contribute to the property of the color 
superconducting phase. Therefore, there is no gluon free-energy in the
gNJL model,  it is not clear how to derive the gluon condensate 
in this model.   In order to investigate the problem in a fully self-consistent 
way, one has to use the ambitious framework by using the Dyson-Schwinger 
equations (DSE) \cite{DSE} including diquark degree of freedom \cite{DSE-SC}
or in the framework of effective theory of high-density quark matter as 
in Ref. \cite{EFT-HDQ}.

\section{Conclusion and discussion}
\label{sec-sum}

In this article, we reviewed the instabilities driven by mismatch and recent
progress of resolving instabilities in the 2SC phase. 

Except the chromomagnetic instability, 
the g2SC phase also exhibits a paramagnetic response to the perturbation 
of an external baryon current. This suggests a baryon current can be spontaneously
generated in the g2SC phase, and the quasiparticles spontaneously obtain 
a superfluid velocity. The spontaneously generated baryon current breaks the 
rotational symmetry of the system, and it resembles the one-plane wave
LOFF state.

We further describe the 2SC phase in the nonlinear realization framework,  and 
show that each instability indicates the spontaneous generation of the corresponding 
pseudo Nambu-Goldstone current. We show this Nambu-Goldstone currents generation
state can naturally cover the gluon phase as well as the one-plane wave LOFF state.

We also point out that, when charge 
neutrality condition is required, there exists a narrow unstable LOFF 
(Us-LOFF) window, where not only off-diagonal gluons but the diagonal 8-th 
gluon cannot avoid  the magnetic instability. The diagonal gluon's
magnetic instability in this Us-LOFF window cannot be cured by off-diagonal 
gluon condensate in color superconducting phase. More interestingly, 
this Us-LOFF window will also show up in some constrained Abelian 
asymmetric superfluid system. 

The Us-LOFF window brings us a new challenge. We need new thoughts
on understanding how a BCS supercondutor will be eventually destroyed 
by increasing the mismatch, we also need to develop new methods to
really resolve the instability problem. Some methods developed in 
unconventional superconductor field, e.g., High-$T_c$ superconductor,
might be helpful. 

Till now, the results on instabilities are based on 
mean-field (MF) approximation. The BCS theory at MF can describe
strongly coherent or rigid superconducting state very well. 
However, as we pointed out in \cite{Huang-PKU},  with the increase of mismatch, 
the low degrees of freedom in the system have been changed. For example, the gapless
quasi-particle excitations in the gapless phase, and the small Meissner mass square of the 
off-diagonal gluons around $\delta\mu=\Delta/\sqrt{2}$. This indicates that these 
quasi-quarks and gluons become low degrees of 
freedom in the system, their fluctuations become more important.  
In order to correctly describe the system, the low degrees of freedom should be 
taken into account properly.  The work toward this direction is still in progress.   
 
In this article, we did not discuss the magnetic instabilty in the gCFL phase. 
For more discussion on solving the magnetic instabilty
in the gCFL phase, please refer to Ref. \cite{gCFL-LOFF}.  

\section*{Acknowledgments}

The author thanks M. Alford, F.A. Bais, K. Fukushima, E. Gubankova, M. Hashimoto, T. Hatsuda, 
L.Y. He, D.K. Hong, W.V. Liu, M. Mannarelli, T. Matsuura, Y. Nambu, K. Rajagopal, H.C. Ren, 
D. Rischke, T. Schafer,  A. Schmitt,  I. Shovkovy, D. T. Son, M. Tachibana, Z.Tesanovic, X. G. Wen, 
Z. Y. Weng, F. Wilczek and K. Yang for valuable discussions. The work is supported by the Japan Society 
for the Promotion of Science fellowship program.

\end{document}